# Comparison of Scatchard and Klotz methods for Metal(II) -DNA complexes in the case two type of binding.

**Eteri Gelagutashvili**

E.L. Andronikashvili Institute of Physics, 6 Tamarashvili str.,

Tbilisi, 0177, Georgia

E.mail: gel@iphac.ge

gelaguta@yahoo.com

## Abstract

It was shown, that modified Scatchard plots and Klotz properties of grafhical representations of two classes of binding sites could be used to determine binding constants of metal ion to DNA. More realistic picture is obtained by Scatchard plots.

The determination of equilibrium binding constants is a common and essential tool for investigation biological macromolecules and ligands biocomplexes of metal.

The development of several different methods suitable for calculating equilibrium constants. Most methods are based on linearization procedures of the law of mass action[1,2]. The improvement of computer technology led to the development of non-linear curve fitting algorithms[3,4]. Scatchard and Klotz method has been widely used to analyze experimental data.

In this article we describe, as an example, various binding models of lead in it divalent oxidation state to DNA. In particular, modified Scatchard Plots and Klotz method and their application to the binding of Pb(II) ions to native DNA is shown.

## Materials and methods

DNA samples was used from calf thymus „merck''. Lead salt (Pb(NO$_3$)$_2$) was reagent grade. Complex formation were investigated at 0.002 M Na(I) and at room temperature. DNA concentrations were determined spectrophotometrically using molar extinction coefficient $\varepsilon_{258\,nm}$ =6600 M$^{-1}$ cm$^{-1}$. Equilibrium dialysis experiments were performed in a two-chambered Plexiglass apparatus. Samples were analyzed for Pb(II) by flame atomic absorption spectrophotometry using a wavelength of 283,3 nm. Pb(II) binding to DNA were analyzed by the graphic methods of Scatchard and Klotz [5,6].

## Results and analysis

Determining the binding constants of metal complexes to DNA is of paramount importance in the development of cleavage agents for probing nucleic acid structure and other applications [7]. Dialysis is the most widely applied separation technique.

**Table 1.** The adsorption isotherm of Pb(II) ion with DNA.

| r | r/m | yt | dye | dyt |
|---|---|---|---|---|
| .015 | .910 | .923 | .073 | -.013 |
| .016 | .880 | .909 | .060 | -.029 |
| .020 | .860 | .854 | .052 | .006 |
| .043 | .690 | .600 | .075 | .090 |
| .060 | .510 | .482 | .071 | .028 |
| .080 | .380 | .394 | .036 | -.014 |
| .083 | .370 | .384 | .038 | -.014 |
| .100 | .320 | .337 | .035 | -.017 |
| .170 | .220 | .229 | .021 | -.009 |
| .220 | .170 | .181 | .015 | -.011 |
| .350 | .100 | .086 | .008 | .014 |
| .420 | .040 | .040 | .002 | .000 |

This technique provide quantitative binding data in terms of free (m) and bound ($c_b$) ions. We describe here models which have been used to analyze binding data for metal complexes with DNA. In Table 1 are presented measured (r and r/m) and calculated (yt) values and dye - experimental standard deviation and dyt = r/m - yt, as Scatchard coordinates, where r is the concentration of bound lead ions, m is the concentration of free lead ions.

Their graphical equivalent are shown in fig.1-(A) and (B). Nonlinearity of the binding plot may arise from several effects which are difficult to distinguish: overlap of binding sites, cooperative effects, existence of two different binding sites. Using of a number of mathematical models reveals that in the case Pb(II) - DNA complexes two independent site binding is the most succesive approximation.

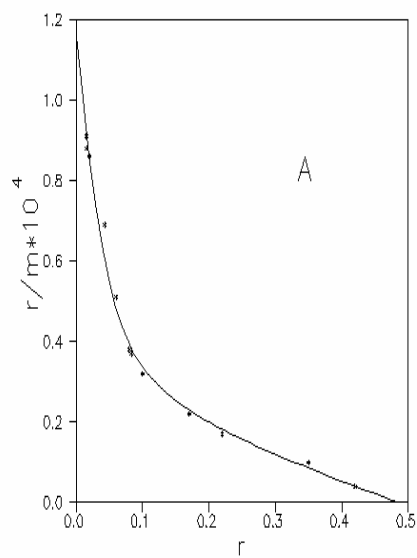

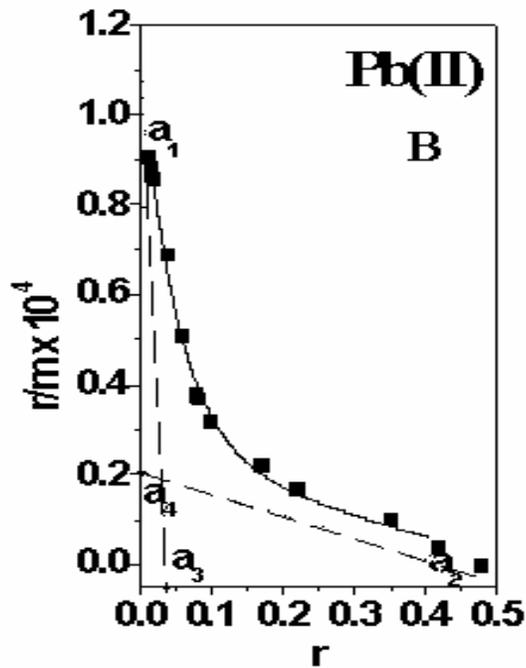

???. 1. – (A) -The Scatchard plots for binding of Pb(II) with DNA (2mM NaNO$_3$, 20°C) as non-linear least square fit to equation:

$r/m = 0.5 [ B(r) + \sqrt{B^2(r) + 4C(r)} ]$             (1)

$C(r) = k_1 k_2 r (n_1 + n_2 - r)$. $r$- is the concentration of bound lead ions, $m$- is the concentration of free lead ions.

$k_1$, $k_2$ are microconstants and $n_1$, $n_2$ - the number of binding sites for lead ions per phosphate group of DNA

(B) – The Klotz plots for binding of Pb(II) with DNA (2mM NaNO$_3$, 20°C).      Parameters $r$ and $m$ are the same as in Fig. 1(A). Calculated values ($k_1, n_1, k_2, n_2$) where      $B(r) = k_1 n_1 + k_2 n_2 - (k_1 + k_2) r$ were obtained from the Klotz equation system: $\begin{cases} a_1 = n_1 k_1 + n_2 k_2 \\ a_2 = n_1 + n_2 \\ a_3 (n_1 k_1^2 + n_2 k_2^2) = a_1^2 \\ (n_1 k_2 + n_2 k_1) a_4 = a_2^2 k_1 k_2 \end{cases}$

The basic binding model of Scatchard derives from a simple consideration of the law of mass action, where n - represents the apparent number of binding sites per P and $K_{app.}$ is the apparent binding constant.

The equilibrium binding of metal ions to independent and identical or binding sites at the molecules DNA (so - called Scatchard Plots) can be written as

$$r = \sum_{i=1}^{N} \frac{k_i n_i m}{1 + k_i m} \quad (1)$$

In the case two type of binding

$$r/m = 0.5 \left[ B(r) + \sqrt{B^2(r) + 4C(r)} \right] \quad (2)$$

where
$$B(r) = k_1 n_1 + k_2 n_2 - (k_1 + k_2) r$$
$$C(r) = k_1 k_2 r (n_1 + n_2 - r).$$

$k_1$, $k_2$ are microconstants and $n_1$, $n_2$ the number of binding sites for lead ions per phosphate group of DNA. The intersection of the curves with the abscissa gives the value $n = n_1 + n_2$. (n is the number of binding sites for lead ion per phosphate group of DNA at saturation). and the intersection of the curves with the ordinate gives the value stoichiometric binding constant (K) x $n = k_1 n_1 + k_2 n_2$.

Definition of $k_1$, $n_1$, $k_2$, $n_2$ magnitudes is performed by fitting of the curve (2) to the experimental points. For this, the minimisation of the folliving magnitude is necessary

$$\chi^2 = \frac{1}{N-M} \sum_{i=1}^{N} \left[ y(r_i, \alpha_k) - (r/m)_i \right]^2 \frac{1}{\sigma_i^2} \quad (3)$$

where $\alpha_k = \{k_1, n_1, k_2, n_2\}$ - is the unity of parameters, $\sigma_i$ - is the root - mean-square deviation of the experimental points, N is the point number, M is the number of parameters. Function (2) dependens on $\alpha_k$ parameters nonlinear, therefore the minimum of expression (3) is found by Monte-Carlo method. For parameters $\alpha_k$ their values are drawn randomely in the physically permisible limit and that $\alpha_k$ is chosen which corresponds the minimum of expression (3).

At the same time we calculate the error matrix (covariance matrix) [8], the diagonal elements of which are the errors of the found parameter values.

$$da_k = (z^{-1})_{kk},$$

where $$z_{jk} = \sum_{i=1}^{N} \frac{G_j G_k}{\sigma_i}, \qquad G_k = \partial y / \partial \alpha_k$$

The magnitude (3) obeys $\chi^2$ distribution .thus its minimum value can by used for fitting criterium . In our case $\chi^2 = 0.8 < 1$ i.e. model (2) is suitable.

The same experiment is carried out by Klotz method fitting of the experimental points is performed with hypothetical function $y = f(r, \alpha_k)$ by above described method. According to Klotz at the intersection points of well fitted curve axes tangents should be drawn

$$a_1 = y(0), \quad y(a_2) = 0, \quad a_3 = -a_1/y'(0), \quad a_4 = -a_2 y'(a_2).$$

and a set of equations should be formed.

$$a_1 = n_1 k_1 + n_2 k_2$$
$$a_2 = n_1 + n_2 \qquad \qquad (4a - 4d)$$
$$a_3(n_1 k_1^2 + n_2 k_2^2) = a_1^2$$
$$(n_1 k_2 + n_2 k_1) a_4 = a_2^2 k_1 k_2$$

We solved the set of equation analitically and calculated the errors.

$$k_1 = (P-D)/2A; \quad k_2 = (P+D)/2A; \quad n_1 = -(a_1 - a_2 k_2)A/D; \quad n_2 = a_2 - n_1,$$

where $A = a_2^2 a_3(a_1 - a_4)$, $P = a_1 a_2(a_1 a_2 - a_3 a_4)$, $Q = a_1^2 a_4 (a_2 - a_3)$, $D = \sqrt{P^2 - 4AQ}$

and calculated the errors.

$$\delta k_1 = k_1 \left( \frac{\delta A}{A} + \frac{\delta P + \delta D}{P+D} \right), \qquad \delta k_2 = k_2 \left( \frac{\delta A}{A} + \frac{\delta P + \delta D}{P+D} \right)$$

$$\delta n_1 = n_1 \left\{ (\delta a_1 + k_2 \delta a_2 + a_2 \delta a_2)/(a_1 - a_2 k_2) + \frac{\delta A}{A} + \frac{\delta D}{D} \right\}, \quad \delta n_2 = \delta a_2 + \delta n_1.$$

Table 2. summarizes all results. It is clear from Table 2, that all parameters determined with eq.2 is almost identical with the values, which were determined by eq. 4a - 4d. The stoichiometric binding constant (K) using eq.2 agree very well with those obtained from eq. 4a - 4d. Consequently, the results in Table 2 suggested to us that both models (modified Scatchard plots and Klotz properties of graphical representations of two classes of binding sites) could by used to determine binding constant of Pb(II) to DNA. But, using eq. 4a - 4d to determine binding constant gives high $\chi 2$, however, a more realistic picture is obtained, when the data fit. to eq.2 (graphical illustration is fig.1).

Analogue results were received for Ni(II) _ and Zn_DNA complexes.

**Table 2.** Parameters for Pb(II) ions binding to DNA in 0.002 M Na(I) at $20^0$C.

|  | Determined with eq.2 | Determined with 4a -4d |
|---|---|---|
| $k_1 \times 10^4$, $M^{-1}$ | 21.07 ±0.69 | 24.68 ± 1.16 |
| $k_2 \times 10^4$, $M^{-1}$ | 0.56 ± 0.01 | 0.68 ± 0.014 |
| $n_1$ | 0.043 ±0.021 | 0.034 ± 0.001 |
| $n_2$ | 0.442 ±0.025 | 0.466 ± 0.001 |
| $K \times 10^4$, $M^{-1}$ | 2.42 ± 0.48 | 2.33 ± 0.158 |
| $\chi^2$ | 0.81 | 3.08 |

It is seen also, from Table 2 that microconstant ($k_1$) for the first site is greater than microconstant ($k_2$) for the second site. Thus, Pb(II) - DNA interaction were observed two - type of binding: strong interaction (with microconstant $k_1$) and weak interaction (microconstant $k_2$). It appears that in this oxidation state lead binds preferentially to thiol and phosphate groups in nucleic acids, proteins and cell membranes [9]. The ``hard`` metal ions Mg(II) and Eu(III) tend to bind preferentialy to the oxygen ligands and phosphates, while the ``soft`` ions Pb(II) and Mn(II) interact rather with the aromatic nitrogen atoms of nucleic acid bases [10].

R E F E R E N C E S